\documentclass[prd,showpacs,showkeys,preprint]{revtex4}

\usepackage{amsmath}
\usepackage{amssymb}
\usepackage{yfonts}[1988/10/03]
\usepackage{graphicx}
\newcommand {\beq}{\begin{equation}}
\newcommand {\eeq}{\end{equation}}
\newcommand {\beqa}{\begin{eqnarray}}
\newcommand {\eeqa}{\end{eqnarray}}

\begin{document}
\title{ Relativistic Bohmian Mechanics } 
\author{Jafar Khodagholizadeh}
\email{j.gholizadeh@modares.ac.ir}
\affiliation{Faculty of Basic Science , Shahid Beheshti branch ,Farhangian University, Tehran, Iran.}
\author{Javad Kazemi}
\email{kazemi.j.m@gmail.com}
\affiliation{ Department of Physics, The University of Qom, Qom, Iran}
\author{Alireza Babazadeh}
\email{arbabazade@yahoo.com} 
\affiliation{  Amirkabir University of Technology, 424 Hafez Ave, Tehran, Iran}

\begin{abstract}
 In this paper we generalize the ideas of de Broglie and Bohm to the relativistic case which is based on the relativistic Schrodinger equation. With this approach, the relativistic forms of the guidance equation and quantum potential are derived. In our formulation of the Relativistic Bohmian Mechanics, the quantum equilibrium hypothesis $(\rho=|\psi|^{2})$ and the probabilistic interpretation of the wave function are not an intrinsic feature of the theory as expected from the theoretical structure of Bohmian mechanics, however we can still extract the statistical predictions of the considering theory. By assuming the quantum equilibrium hypothesis in the non-relativistic case and acceleration of particles by an external field, we go to the relativistic regime. In this case the quantum equilibrium would not be established and the theory will have testable predictions that can be compared to the results of relativistic quantum mechanics. Here we present the calculations for some simple examples.
\keywords{Bohmian Mechanics,  Relativistic Quantum Mechanics}

\end{abstract}
\maketitle

\section{Introduction}
One of the basic concepts of quantum mechanics widdly accepted in the scientific culture, is that natural processes are inherently non-causal\cite{Weinberg},although some of the contemporary physicists such as Lorentz, de Broglie, Schrodinger, Einstein, Bohm and Bell disagreed with this view \cite{Valentini}-\cite{Holland}. Within this context in 1952 David Bohm presented a causal theory to describe quantum systems which is known Bohmian mechanics\cite{Bohm}-\cite{Bohm3}. Beside the Copenhagen formulation this theory became also widespread and reproduced all statistical result of quantum mechanics. At first Bohmian mechanics was formulated by de Broglie between 1923 and 1927\cite{de Broglie} and then in 1952 Bohm completed it independently and answered many questions related to the theory of de Broglie\cite{Bohm3}-\cite{Durr}. Despite its nonrelativistic nature,  Bohmian mechanics has been very successful and consistent but there are also some technical and principal problems. Some attempts have been made to formulate its relativistic generation \cite{Holland} but most of these efforts have not been very successful\cite{Holland} ,\cite{peter}-\cite{Nicoli'c4}. In this paper we introduce a new relativistic derivation of the Bohmian mechanics that enables us to go beyond the predictions of the standard quantum mechanics which is empirically testable.
At first, we give a brief review of the concepts and formalisms of nonrelativistic Bohmian Mechanics and then we develop the relativistic case.
\section{Non Relativistic Bohmian Mechanics }
     In the de Broglie-Bohm mechanics, a quantum system is formed from particles in addition to waves and particles move in well-defined trajectories. In this theory, quantum effects are explained by the aid of a “quantum force” which is  exerted on a particle from the $ \psi $ field and using this coupled dynamics, some wave-like behaviors are transferred from the $\psi$ field to the particle.
     In Bohmian Mechanics probability is not considered as an intrinsic property but the stochastic characteristics of quantum effects are explained by the aid of initial stochastic conditions \cite{Bohm3}-\cite{Durr}. 
The de Broglie-Bohm mechanics gives a thorough explanation for all quantum phenomena and even some phenomena which seem to be queer. Atomic transitions and measurement processes in the Copenhagen interpretation will have very simple and comprehensible explanations in de Broglie-Bohm formulation\cite{Holland}, \cite{Durr0}.  In order to have a better comprehension of this model,  the following explanation could be helpful.

     Bohm derived equations by replacing the polar form of the wave function $\psi=R e^{\frac{iS}{\hbar}}$  within the Schrodinger equation, and separating the real and imaginary parts:
\begin{equation}\label{1}
\frac{\partial R^{2}}{\partial t}+\vec{\nabla}.(R^{2}\frac{\vec{\nabla}S} {m})=0 
\end{equation}
\begin{equation}\label{2}
\frac{\partial S}{\partial t}+\frac{1}{2m}(\vec{\nabla}S)^{2}+U(x)-\frac{\hbar^{2}}{2m}\frac{\nabla^{2}R}{R}=0
\end{equation}
Bohm considered eq.(\ref{1}) as the continuity equation and eq.(\ref{2}) as a generalized Hamilton-Jacobi equation. The last term in eq. (\ref{2}) is the same term which was called as Quantum Potential by Bohm called $Q$ . We also recall that in the Hamilton-Jacobi formulation, the action gradient $S$ , is equal to the particle momentum $\vec{P}=\vec{\nabla} S$. With regard to the relationship between velocity and momentum in nonrelativistic mechanics we have
\begin{equation}\label{3}
\vec{v}=\frac{\vec{\nabla} S}{m}
\end{equation}
In addition, eq. (\ref{3}) is in harmony with the interpretation of eq. (\ref{1}) as the continuity equation. These simple equations made just an adequate theme for Bohm to present a deterministic interpretation of quantum mechanics. Bohm assumed that  the phase of the wave function would give us the particle’s velocity, in accordance to the eq. (\ref{3}). Thus by having the equation for the time revolution of the phase of the wave function (the Schrödinger’s equation), the velocity of the particle can be determined at every moment and calculate its trajectory using the initial location.  Therefore Bohm considered this a causal dynamics for the movement of particles through realistic trajectories. In order to compare this theory to classical mechanics, one ought to calculate the acceleration of particles in its “Bohmian trajectories” :
\begin{equation}
m \frac{d \vec{v}}{dt}=m(\frac{\partial}{\partial t}+\vec{v}.\vec{\nabla})\vec{v}=\vec{\nabla}\frac{\partial S}{\partial t}+(\vec{\nabla}S.\vec{\nabla})\vec{\nabla}S
\end{equation}
Using eq. (\ref{2}) :
\begin{equation}
m \frac{d \vec{v}}{dt}=-\vec{\nabla}(U+Q)
\end{equation}
As we can see, particles in Bohmian trajectories are influenced by a quantum potential in addition to the classical potential. Bohm considered this quantum potential as the difference between quantum mechanics and classical mechanics. After a short introduction we will review theoretical structure and the mathematical formalism of non-relativistic Bohmian mechanics considering more details in the next section.
\subsection{Bohmian Mechanics: The minimal approach}
     In order to formulate the Bohmian mechanics axiomatically, we use an approach known as “the minimal approach”. This approach has its roots in de Broglie’s works and some further developments by some physicists like Zanghi, Goldstein and Durr\cite{Berndl}-\cite{Durr3}. In this approach, the state of a quantum system is determined by the pair $(\vec{X},\psi)$. In which $\psi$ is the wave function and $\vec{X}$ determines the location of the particle. In this formalism, two first order differential equations (as axioms) are assumed ; one for the time evolution of the particle’s location and another for the wave function \cite{Berndl}. Regarding the polar form of the wave function,  $\psi=R e^{\frac{iS}{\hbar}}$, the time evolution equations are given below:
\subsubsection{First Axiom: The equation of motion of a particle (guidance equation)}
    The following equation which is known as the “guidance equation ” is considered as the equation of motion of a particle:
\begin{equation}\label{6}
 \frac{d \vec{X}(t)}{dt}=\frac{\vec{\nabla} S}{m}\mid_{ x=\vec{X(t)}}
\end{equation}
Although it is common that in de Broglie-Bohm formulation, eq. (\ref{6}) is considered as an independent axiom, some aouthers tried to derive it from some predicates such as Galilean Invariance, time reversibility and simplicity \cite{Durr3}. It is indeed quite clear that regarding the Hamilton-Jacobi formulation for describing motion of the particles, the equation $\vec{P}=\vec{\nabla} S$ is trivial.

\subsubsection{Second Axiom: The time revolution equation of the wave function}
In Bohmian mechanics as well as standard quantum mechanics, Schrödinger ’s equation is considered for the time revolution of the wave function: 
\begin{equation}
 i\hbar \frac{\partial \psi}{\partial t}=\frac{-\hbar^{2}}{2m}\nabla^{2}\psi + U(x)\psi
\end{equation}
The first and second axioms establish a deterministic dynamics for motion of a particle. Of course careful attention is needed to the fact that by knowing the exact initial location of the particle, its location in all future will be determined uniquely, and by knowing the initial location of the particle statistically, the further locations of the system will be known to us statistically. 
\subsubsection{Third Axiom: Quantum Equilibrium}
The statistical distribution of the initial location of particles is considered as follows:
\begin{equation}\label{8}
\rho_{0}=|\psi_{0}|^{2}
\end{equation}
where $\rho_{0}$ is initial probability density of the particle and $\psi_{0}$ is the initial wave function of the system.
Their assumption that eq.(\ref{8})applies, is known as the quantum equilibrium. Attention is needed here that by assuming eq. (\ref{8}) in one moment, will also applies in the furthertimes, which is a result of the first and second axioms. In fact, the first and second axioms direct the particles in trajectories in which $|\psi|^{2}$   is preserved as the probability density as time passes. This issue causes consistency between the statistical results of this theory and the standard quantum mechanics \cite{Holland}, \cite{Durr}. 
At first this property for the distribution function $|\psi|^{2}$ was introduced by de Broglie in 1927 and was later explicitly formulated by Durr and called equivariance\cite{Durr3}.
It has been shown that the only local function of $\psi$  which has the  equivariance property is $|\psi|^{2}$  \cite{Goldstein}.

Of course, some efforts have been made to show that the third axiom results from simple assumptions. As an example, the relaxation of an arbitrary distribution function $\rho$  to the quantum equilibrium distribution $(\rho \longrightarrow |\psi|^{2})$ has been studied by several researchers \cite{Antony}-\cite{Colin2}. Some authors have also considered the explanation of this distribution functions as the concept of  "typicality " \cite{Berndl}, \cite{Durr3}. Such studies have somewhat justified the concept of " quantum equilibrium" .

Regarding what was remarked so far, we should say that the probability interpretation of the wave function is not an intrinsic part of Bohmian theory and this theory is completely sensible even without such a concept.  In fact in Bohmean theory,  $\psi$ is considered as a real field which has as influence upon the dynamics of particles and the relation between the wave function and the location probability density plays only a secondary role in wave function. The assumption of the quantum equilibrium is only added in order to produce results similar to the standard quantum mechanics.
 \section{Relativistic Generation of de Broglie-Bohm Mechanics}
 Bohmian theory was quite successful at non-relativistic but the attempt to extend the causal interpretation to relativistic quantum mechanics are not successful. As we know relativistic particles with zero spin are described by the Klein-Gordon equation but there is an incompatibility of the Klein-Gordon equation with a probabilistic interpretation of the wave function. Therefore describing particles in the framework of Bohmian mechanics as well as standard quantum mechanics is an essential problem, similarly defining the velocity field (guidance equation) based on the Klein-Gordon equation is a fundamental problem\cite{Holland}.  Despite these problems, efforts have been made to generalize relativistic Bohmian mechanics and some of these efforts have been partially successful[\cite{peter},\cite{Fathimah}-\cite{Nicoli'c4}] but not satisfactory enough. In this paper we present a simple relativistic generalization of the de Broglie-Bohm theory based on the relativistic Schrodinger equations.
 \subsection{Postulate of relativistic Bohemian mechanics }
 We consider a minimalist approach for the formulation of the relativistic Bohmian mechanics  as a non-relativistic theory with an equation for the time evolution of the wave function and an equation for the time evolution of the particle position which is affected by the wave function.
 \subsubsection{Time evolution equation of the wave function}
 We consider the relativistic Schrödinger equation for the time evolution of the wave function:
 \begin{equation}\label{9}
 i\hbar \frac{\partial \psi}{\partial t}=\sum_{k=0}^{\infty}a_{k}\hbar^{2k}\nabla^{2k}\psi
 \end{equation}
 The above equation is obtained from the relativistic dispersion relation with $ E= i\hbar \partial /\partial t $  and $ \vec{p} = -i\hbar \nabla $ :
 \begin{eqnarray}\label{10}
 E=\sqrt{c^{2}\vec{p}^{2}+m_{0}^{2}c^{4}}=\sum_{k=0}^{\infty}a_{k} p^{2k} :
 \end{eqnarray}
 where $ a_{k}=(m_{0}c^{2}(-1)^{k} \frac{1}{2k-1}\frac{1}{2^{2k}}\frac{(2k)!}{k!}$.  If the series does not converge we can  use a form of the integral-differential of the relativistic Schrodinger equation :
 \begin{equation}\label{11}
 i\hbar \frac{\partial \psi}{\partial t}=\int_{-\infty}^{+\infty} K(x-x^{'}) \psi(x^{'})dx^{'}
 \end{equation}
 where
 \begin{equation}
 K(x)=\frac{N_{1}(i x/l_{0})-N_{1}(-i x/l_{0})}{4i x/l_{0}}
 \end{equation}
 and $ N_{1} $ is the Neumann function. It is easy to show that if the Eq.(\ref{10}) is convergent  the Eq.(\ref{9}) is equivalent to Eq.(\ref{11}). Also Eq.(\ref{11}) is derived individually from the following assumptions :
 
 $\bigstar $  The equation is linear.
 
 $\bigstar $ The equation is of first order with respect to time.
 
  $\bigstar $ The equation leads to the relativistic dispersion relation for plane wave:
 \begin{eqnarray}
 \omega(k)=\sqrt{c^{2}(\hbar k)^{2}+m_{0}^{2}c^{4}}
 \end{eqnarray}
 where $ \omega $ is a frequency of the wave function and $ k $ is the wave number.
 \subsubsection{The equation of motion of particles}
 For the formulation of the relativistic Bohmian mechanics we consider $ \vec{p}=\vec{\nabla}S$ as the law of particle motion where  $ S $ is the action in Hamilton-Jacobi theory and the relativistic relation between momentum and velocity is $ \vec{p}=\frac{m_{0}\vec{v}}{\sqrt{1-\frac{v^{2}}{c^{2}}}} $,  so the relativistic equation of guidance will be 
 \begin{equation}\label{14}
 \vec{v}_{re}=\frac{c^{2}\vec{\nabla}S}{\sqrt{(\vec{\nabla}S)^{2}c^{2}+m_{0}^{2}c^{4}}}
 \end{equation}
 the above relation is the relativistic generalization of de Broglie guidance equation and it becomes in the first approximation the usual nonrelativistic guidance equation :
 \begin{eqnarray}
 |\frac{\vec{\nabla}S}{m_{0} c}|\ll 1 \Longrightarrow v\approx \frac{\vec{\nabla}S}{m_{0}}
 \end{eqnarray} 
 Equations (\ref{9}) and (\ref{14}) are the basic principles of the theory which can be established as a causal dynamics for particles motion. In the following with some example we will discuss the content of the theory and show its relationship with the corresponding nonrelativistic cases.
 \subsubsection{Example: Plane Waves}
 At first we investigate a plane wave:
 \begin{equation}
 \psi(\vec{x},t)=A e^{i(\vec{k}.\vec{x}-\omega_{k}t)}
 \end{equation}
 where $ \omega_{k}=\frac{1}{\hbar} \sqrt{( \hbar k)^{2}c^{2}+m_{0}c^{4}}$  is a the frequency of the wave function and $ k $ is the wave number. Energy and momentum of the particles are usual given by:
 \begin{eqnarray}
 \vec{p}=\vec{\nabla}S=\hbar\vec{k}
 \end{eqnarray}
 and
 \begin{eqnarray}
 E:=-\frac{\partial S}{\partial t}=\sqrt{( \hbar k)^{2}c^{2}+m_{0}^{2}c^{4}}
 \end{eqnarray}
  from Eq.(\ref{14}) the particle trajectory is
  \begin{eqnarray}
  \vec{x}(t)=\frac{\hbar \vec{k}c^{2}}{\sqrt{(\hbar \vec{k})^{2}c^{2}+m_{0}^{2}c^{4}}}t+\vec{x}_{0}
  \end{eqnarray}
  Thus particles move uniformly and their speed are less than the speed of the light.
  
  To investigate the superposition of plane waves, we consider the wave function of a system as a superposition of two wave functions:
  \begin{eqnarray}
  \psi(\vec{x},t)=A [\psi _{1}(\vec{x},t)+b\psi_{2}(\vec{x},t)] 
  \end{eqnarray}
 which $ \psi_{j}(\vec{x},t) $ is
  \begin{eqnarray}
  \psi_{j}(\vec{x},t)=e^{i(\vec{K}_{j}.\vec{x}-\omega_{k_{j}}.t)}~~~~~,~~~~~ j=1,2
  \end{eqnarray} 
  where  $ \omega_{\vec{k}_{j}}=\frac{1}{\hbar}\sqrt{( \hbar \vec{k}_{j})^{2}c^{2}+m_{0}^{2}c^{4}} $:
  \begin{eqnarray}
  A=\mid A\mid e^{i\varphi}~~~~~~~,~~~~~~~b=\mid b\mid e^{i\delta}
  \end{eqnarray}
  $ A $ and $ b $  are complex constants that $ |b| $ determines the relative amplitude of the waves.
  Using Eq.(\ref{14}), the relativistic equation of particles motion is obtained as follows:
  \begin{eqnarray}
  \frac{d x(t)}{dt}=\frac{\frac{\hbar c^{2}(\vec{k}_{1}+\vert b\vert^{2}\vec{k}_{2}+\vert b\vert (\vec{k}_{1}+\vec{k}_{2})\cos \xi)}{1+\vert b\vert^{2} +2\vert b\vert \cos \xi}}{\sqrt{c^{2}(\frac{\hbar(\vec{k}_{1}+\vert b\vert^{2}\vec{k}_{2}+\vert b\vert (\vec{k}_{1}+\vec{k}_{2})\cos \xi)}{1+\vert b\vert^{2} +2\vert b\vert \cos \xi})^{2}+m_{0}c^{4}}}\vert_{x=x(t)}
  \end{eqnarray}
 where
 \begin{equation}
 \xi \equiv (\vec{k}_{2}-\vec{k}_{1}).\vec{x}-(\omega _{k_{2}}-\omega_{k_{1}})t-\delta
 \end{equation}
 is the relative phase. In the nonrelativistic case, the equation of motion is 
 \begin{eqnarray}\label{25}
 \frac{d x_{nr}(t)}{dt}=\frac{\hbar}{m_{0}}\frac{\vec{k}_{1}+\vert b\vert^{2}\vec{k}_{2}+\vert b\vert (\vec{k}_{1}+\vec{k}_{2})\cos \xi_{nr}}{1+\vert b\vert^{2} +2\vert b\vert \cos \xi_{nr}}
 \end{eqnarray}
 where $ \xi_{nr} $ is
 \begin{eqnarray}
 \xi_{nr} \equiv (\vec{k}_{2}-\vec{k}_{1}).\vec{x}-(\frac{\hbar \vec{k}_{2}^{2}}{2m}-\frac{\hbar \vec{k}_{1}^{2}}{2m})t-\delta
 \end{eqnarray}
 as expected when $ \hbar k_{1,2}\ll m_{0}c$ i.e.  the Compton  wavelength is much shorter than the de Broglie wavelength and the relativistic equation of motion becomes the nonrelativistic eq.(\ref{25}.
 \subsubsection{Example: Particle in the Box} 
 We consider a particle in this potential:
 \begin{eqnarray}
 v(x)=\lbrace^{\infty ~~~~~~ x<0 ~and~x>L}_{0~~~~~~0\leq x\leq L}
 \end{eqnarray}
 The eigenstates and eigenvalues of the system are obtained from the equation $ (\sum_{k=0}^{\infty}a_{k}\frac{\partial^{2}}{\partial x^{2k}})\psi_{n}(x)=E_{n}\psi_{n}(x) $ with the boundary condition $ \psi(L)=\psi(0)=0$ :
 \begin{eqnarray}\label{29}
 E_{n}=\sqrt{(\frac{n\pi \hbar}{L})^{2}c^{2}+m_{0}^{2}c^{4}}
 \end{eqnarray}
 and
 \begin{eqnarray}
 \psi_{n}(L)=A_{n}\sin(\frac{2n \pi x}{L})
 \end{eqnarray}
 The eigenstates are identical with the non-relativistic case but the relativistic eigenvalues differ from the nonrelativistic eigenvalues and as expected if the size of the potential well is large compared to the Compton wavelength, the nonrelativistic eigenvalue derived with extension of (\ref{29}) to a the first order approximation is:
 \begin{eqnarray}
 \frac{l_{c}}{L}\ll 1~~~~~\longrightarrow ~~~~~E_{n}\approx m_{0}c^{2}+\frac{n^{2}\pi^{2}\hbar^{2}}{2m_{0}L^{2}}
 \end{eqnarray}
 where $ l_{c}\equiv \frac{\hbar}{m_{0}c} $ is the Compton wavelength. As it is obviously, stationary states(energy eigenfunction) are real so the Bohmian velocity of particles would be zero. To evalute the Bohmian trajectories we consider the superposition of the ground state and the first excited state :
 \begin{eqnarray}
 \psi(x,t)=\psi_{1}(x)e^{-\frac{iE_{1}t}{\hbar}}+\psi_{2}(x)e^{-\frac{iE_{2}t}{\hbar}}
 \end{eqnarray}
 It can be easily shown that the relativistic guidance equation (\ref{14}) with condition $ l=\frac{L}{l_{c}}$ becomes the nonrelativistic equation(6). In figure 1. the relativistic and nonrelativistic trajectories of particles for different values of dimensionless quantity $ l=\frac{L}{l_{c}} $ with initial position $ x_{0}=\frac{L}{2} $ are drawn. When the length of the box, e.g.$ L $ is small compared to the Compton wavelength ,relativistic and nonrelativistic trajectories are separated and when $ L $ is much larger than the Compton wavelength, the relativistic trajectories are consistent with the nonrelativistic case.
 \begin{figure}
 \includegraphics[scale=0.5]{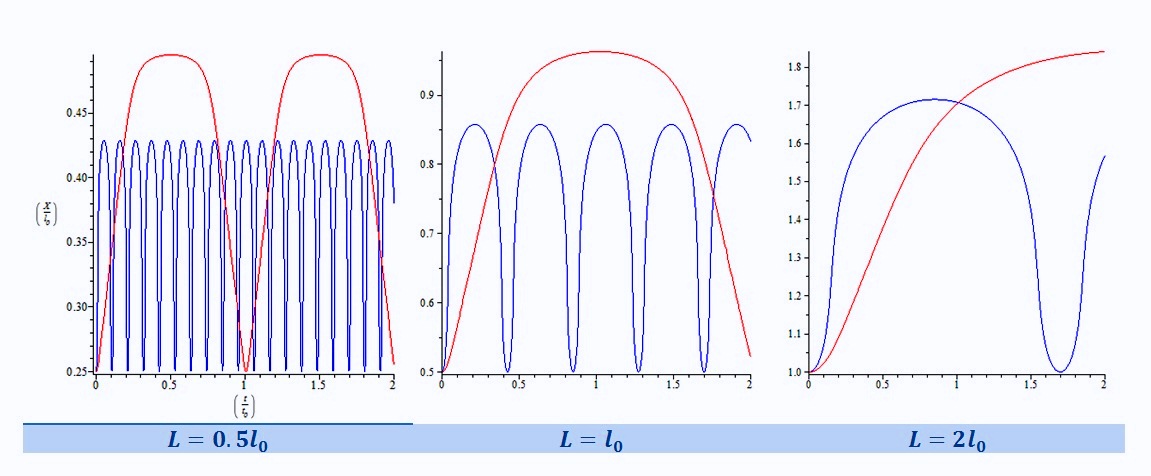}
 \includegraphics[scale=0.5]{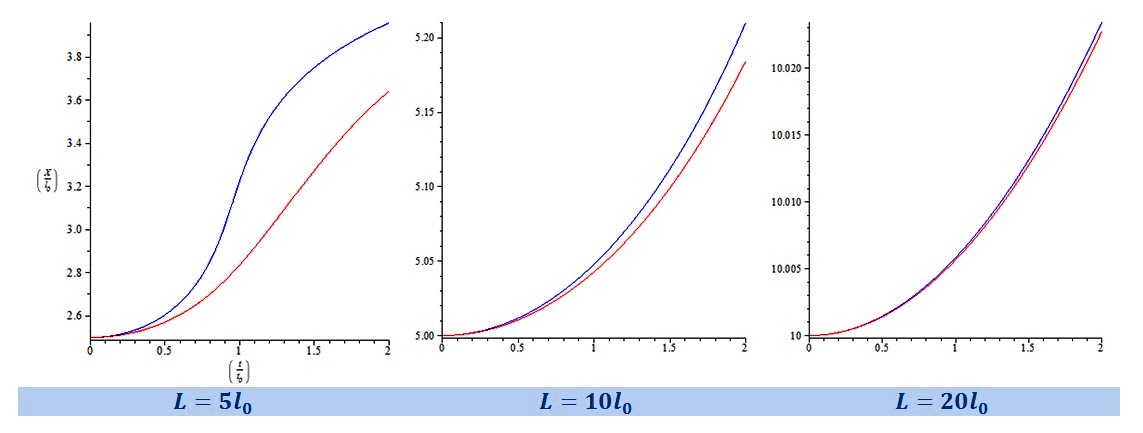}
 \caption{relativistic Bohemian trajectories (red lines) and nonrelativistic (blue lines) for different values  of the box  length (L) are plotted. At the limit $ \frac{L}{l_{c}}\longrightarrow \infty  $ relativistic trajectories are consistent with nonrelativistic trajectories. }
 \label{fig:1} 
 \end{figure}
 \subsection{Force and  the Relativistic Quantum Potential}
 In this section we want to calculate the relativistic formulation of the quantum force according to definition. We have 
 \begin{eqnarray}
 \vec{F}=\frac{d\vec{p}}{dt}=(\frac{\partial}{\partial t}+\vec{v}.\vec{\nabla}\vec){p}
 \end{eqnarray}
 from $ \vec{p}=\vec{\nabla}S$ and  eq.(\ref{14}) and we have 
 \begin{eqnarray}
 \vec{F}=\nabla(\frac{\partial S}{\partial t})+(\frac{c^{2}}{\sqrt{(\vec{\nabla}S)^{2}c^{2}+m_{0}^{2}c^{4}}})(\vec{\nabla }S.\vec{\nabla})\vec{\nabla}S
 \end{eqnarray}
 substituting $ \psi =R e^{i\frac{S}{\hbar}} $in the relativistic Schrödinger equation and separating the real and the imaginary parts we have 
 \begin{eqnarray}
 \frac{\partial}{\partial t}+\eta=0
 \end{eqnarray}
 \begin{eqnarray}\label{24}
 \frac{\partial S}{\partial t}+\zeta=0
 \end{eqnarray}
 where $ \eta  $ and $ \zeta $ are
 \begin{eqnarray}
 \eta\equiv\frac{-1}{\hbar}\sum_{n=0}^{\infty}a_{n}Im[\frac{\vec{\nabla}^{2n}(R \exp \frac{iS}{\hbar})}{\exp \frac{iS}{\hbar}}]
 \end{eqnarray}
 \begin{eqnarray}
 \zeta \equiv \sum_{n=0}^{\infty}a_{n}Re[\frac{\vec{\nabla}^{2n}(R \exp \frac{iS}{\hbar})}{R\exp \frac{iS}{\hbar}}]
 \end{eqnarray}
 and therefor the force can be written as follow
 \begin{eqnarray}
 \vec{F}=-\vec{\nabla}(\zeta)+(\frac{c^{2}}{\sqrt{(\vec{\nabla}S)^{2}c^{2}+m_{0}^{2}c^{4}}})(\vec{\nabla }S.\vec{\nabla})\vec{\nabla}S
 \end{eqnarray}
 It can easily shown that the curl of the second term of the right hand side of above equation is zero, then it can be defined as the gradiant of a function $ \xi $ :
 \begin{eqnarray}
 (\frac{c^{2}}{\sqrt{(\vec{\nabla}S)^{2}c^{2}+m_{0}^{2}c^{4}}})(\vec{\nabla }S.\vec{\nabla})\vec{\nabla}S=-\vec{\nabla}(\xi)
 \end{eqnarray}
 where $ \xi $ is 
 \begin{eqnarray}
 \xi=-\sqrt{(\vec{\nabla}S)^{2}c^{2}+m_{0}^{2}c^{4}}
 \end{eqnarray}
 We can rewrite the quantum force based on the quantum potential
 \begin{eqnarray}
 \vec{F}=-\vec{\nabla}(Q_{re})
 \end{eqnarray}
 in the above relation, the relativistic quantum potential $ Q_{re} $  is defined as :
 \begin{eqnarray}
 Q_{re}\equiv \zeta+\xi
 \end{eqnarray}
 Therefore the relativistic quantum potential is a functional of $ R $ and $ S $:
 \begin{eqnarray}
 Q_{re}[R,S]= \sum_{n=0}^{\infty}a_{n}Re[\frac{\vec{\nabla}^{2n}(R \exp \frac{iS}{\hbar})}{R\exp \frac{iS}{\hbar}}]-\sqrt{(\vec{\nabla}S)^{2}c^{2}+m_{0}^{2}c^{4}}
 \end{eqnarray}
 By substitution  of $ \zeta $ in eq.(\ref{24}) we have
 \begin{eqnarray}\label{45}
 \frac{\partial S}{\partial t}+(\sqrt{(\vec{\nabla}S)^{2}c^{2}+m_{0}^{2}c^{4}})+Q_{re}=0
 \end{eqnarray}
The equation above is the relativistic form of the Hamilton-Jacobi equation that has been modified by the relativistic quantum potential. In order to see that $ Q_{re} $ has the role of an energy, it is better to consider the stationary state (eigenstate of energy) which is  the solution of time independent relativistic Schrödinger equation i.e. $ \hat{H}_{RE}\psi =E\psi $ and $ \psi(\vec{r},t) $ is 
 \begin{eqnarray}
 \psi(\vec{r},t)=\psi_{0}(\vec{r})e^{\frac{-iEt}{\hbar}}
 \end{eqnarray}
 so
 \begin{eqnarray}\label{46}
 \frac{\partial S}{\partial t}=-E ~~~~~~,~~~~~~~\frac{\partial R}{\partial t}=0
 \end{eqnarray}
 By substituting eq.(\ref{46}) in (\ref{45}) we have 
 \begin{eqnarray}
 (\sqrt{(\vec{\nabla}S)^{2}c^{2}+m_{0}^{2}c^{4}})+Q_{re}=E
 \end{eqnarray}
 The statement  above shows that the total energy of the particle is equal to the sum of the kinetic energy (relativistic) and relativistic quantum potential energy, $ Q_{re} $. The relativistic quantum potential introduced in this paper has a similar role as a nonrelativistic quantum potential in the Bohmian theory .
 
 Note that in the stationary state according to eq.(\ref{46}), the quantum potential is time independent ($ i.e.\frac{\partial Q_{re}}{\partial t}=0 $) and the total energy of particles in Bohmian trajectories  is conserved. Furthermore the quantity E in (\ref{46}) is constant and independent of time and position so all particles of the ensemble  have equal energy E. For more clarification lets consider a very simple example.
 \subsubsection{Particle in Box}
 We consider a particle in an infinite potential well( 3-dimensional cube)
 \begin{eqnarray}
 V(x)=\lbrace^{0~~~~~~~~0\leq x,y,z\leq L}_{\infty ~~~~~~~~ other}
 \end{eqnarray}
In this case the eigenstate and eigenvalue of energy using the relativistic Schrödinger equation will be as 
 \begin{eqnarray}
 E_{n}=\sqrt{[(\frac{n_{x}\pi \hbar}{L})^{2}+(\frac{n_{y}\pi \hbar}{L})^{2}+(\frac{n_{z}\pi \hbar}{L})^{2}]c^{2}+m_{0}^{2}c^{4}}
 \end{eqnarray}
 and
 \begin{eqnarray}\label{50}
 \psi_{n}(x,y,z)=A_{n}\sin(\frac{n_{x}\pi \hbar}{L})\sin(\frac{n_{y}\pi \hbar}{L})\sin(\frac{n_{z}\pi \hbar}{L})
 \end{eqnarray}
 In the  expression above $ n=(n_{x},n_{y},n_{z}) $ has been considered. Whenever as in this example, the wave function of the system is  real, the quantum potential takes the simple form
 \begin{eqnarray}\label{51}
 Q_{re}=\sum_{k=0}^{\infty}a_{k}\hbar^{2k}\frac{\nabla^{2k}R}{R}
 \end{eqnarray}
 The quantum potential above has been extracted by Atigh et al. (\cite{Atigh}) by using a variational principle. After substituting (\ref{50}) in (\ref{51}) we have 
 \begin{eqnarray}\label{52}
 Q_{re}=\sum_{k=0}^{\infty}a_{k}\hbar^{2k}[(\frac{n_{x}\pi \hbar}{L})^{2}+(\frac{n_{y}\pi \hbar}{L})^{2}+(\frac{n_{z}\pi \hbar}{L})^{2}]^{k}-m_{0}c^{2}
 \end{eqnarray}
 in this case $ \vec{\nabla}S=0 $, so Bohm velocity and kinetic energy of the particle is zero and the total energy of the particle will be only the quantum potential energy. Therefore from series (\ref{52}) we have
 \begin{eqnarray}
 E_{n}=m_{0}c^{2}+Q_{re}=\sqrt{[(\frac{n_{x}\pi \hbar}{L})^{2}+(\frac{n_{y}\pi \hbar}{L})^{2}+(\frac{n_{z}\pi \hbar}{L})^{2}]c^{2}+m_{0}^{2}c^{4}}
 \end{eqnarray}
 where the result is the same as the energy of the mode number of n.
 \subsubsection{Weak relativistic regime and the approximate formulation}
 The formulation of Bohmian mechanics introduced in the previous section is extracted from relativistic energy-momentum dispersion equation,i.e.  $ E=\sqrt{c^{2}(\vec{p})^{2}+m_{0}^{2}c^{4}} $.  The relativistic Schrodinger equation is derived from the quantization of this equation:
 \begin{eqnarray}
 E=\sum_{k=0}^{\infty}a_{k}\vec{p}^{2k}
 \end{eqnarray}
 with $ \vec{p}\longrightarrow -i\hbar \vec{\nabla} $ and $ E\longrightarrow i\hbar \frac{\partial}{\partial t}$ we have 
 \begin{eqnarray}\label{54}
 i\hbar \frac{\partial \psi}{\partial t}=\sum_{k=0}^{\infty}a_{k}\hbar^{2k}\vec{\nabla}^{2k}\psi
 \end{eqnarray}
 the relation between the particle velocity and momentum can be derived from the dispersion equation:
 \begin{eqnarray}\label{55}
 \vec{v}=\frac{\partial E}{\partial \vec{p}}=(\sum_{k=0}^{\infty}2k a_{k}\vec{p}^{2k-2})\vec{p}=\sum_{k=0}^{\infty}(2k a_{k}(\vec{\nabla}S)^{2k-2})\vec{\nabla}S
 \end{eqnarray}
 Equations (\ref{54}) and (\ref{55}) with respect to $ N=1 $ lead to the nonrelativistic Bohm theory and choosing $ N=\infty $ leads to the relativistic theory described in the previous section. Between these two extremes ,ie  $ 1<N<\infty $ an approximate formulation for weakly relativistic regime can be given. For example the first modification of nonrelativistic theory taking $ N=2 $ is given by
 \begin{eqnarray}\label{56}
 i\hbar\frac{ \partial \psi^{(2)}}{\partial t}=m_{0}c^{2}\psi^{(2)}-\frac{\hbar^{2}}{2m_{0}}\vec{\nabla}^{2}\psi^{(2)}+\frac{\hbar^{4}}{8m_{0}^{3}c^{2}}\vec{\nabla}^{4}\psi^{(2)}
 \end{eqnarray}
 and
 \begin{eqnarray}
 \vec{v}^{2}=(1-\frac{1}{2}(\frac{\vec{\nabla}S^{(2)}}{m_{0}c})^{2})\frac{\vec{\nabla}S^{(2)}}{m_{0}}
 \end{eqnarray}
 index 2 in the $ \psi^{(2)}, Q^{(2)} $and $ \vec{v}^{(2)} $ point out that these quantities are written in the approximation $ N=2 $.
 
 At this stage,  to obtain a consistent theory we use the same order of approximation, both in the time evolution equation of the wave function and in the equation of particle motion. In this approximation by using the polar form of the wave function in equation (\ref{56}) and separation of the real and imaginary parts leads to the following equations
 \begin{eqnarray}
 \frac{\partial R}{\partial t}+\frac{-1}{\hbar}\sum_{n=0}^{n=2}a_{n}Im[\frac{\vec{\nabla}^{2n}(R~exp \frac{iS}{\hbar})}{exp \frac{iS}{\hbar}}]=0
 \end{eqnarray}
 \begin{eqnarray}
 \frac{\partial S}{\partial t}+\sum_{n=0}^{n=2}a_{n} Re[\frac{\vec{\nabla}^{2n}(R~exp \frac{iS}{\hbar})}{exp \frac{iS}{\hbar}}]=0
 \end{eqnarray}
The equation potential is given by 
 \begin{eqnarray}
 Q^{(2)}[R,S]=\sum_{k=0}^{N=2}a_{k}Re[\frac{\vec{\nabla}^{2k}(R~exp \frac{iS}{\hbar})}{exp \frac{iS}{\hbar}}]-(2a_{k}k(\vec{\nabla}S)^{2k-2})\vec{\nabla}S
 \end{eqnarray}
 \section{Time evolution of the probability density (beyond quantum equilibrium) }
 Up to now we have considered a realistic trajectories for particles motion i.e, eq(1) and (14), then the time evolution of the particles density can be calculated in the relativistic regime via the continuity equation:
 \begin{eqnarray}\label{61}
 \frac{\partial \rho}{\partial t}+\vec{\nabla}.(\rho\vec{v}_{re})=0
 \end{eqnarray}
which takes the form:
 \begin{eqnarray}
 \frac{\partial \rho(\vec{r},t)}{\partial t}+\vec{\nabla}.(\rho \frac{c^{2}\vec{\nabla}S}{\sqrt{(\vec{\nabla}S)^{2}c^{2}+m_{0}^{2}c^{4}}})=0
 \end{eqnarray}
 It should be noted that the continuity equation is not derived from the wave equation but also it is the result of our realistic picture of the particles motion and we have not considered any assumption about the relationship between $ \rho $ and $ \psi $ as quantum equilibrium distribution namely it is proportionally that $ \rho $ is not equal to $ \vert\psi\vert^{2} $. According to the dynamics of the wave function and particle (equations (\ref{14}) and(\ref{9})), the assumption $ \rho=\vert\psi\vert^{2} $ is not a homogeneous condition that be considered as a quantum equilibrium because the relativistic Schrodinger equation leads to the following equation for the time evolution of $ \vert\psi\vert^{2} $.
 \begin{eqnarray}
 \frac{\partial  \vert \psi \vert^{2} }{\partial t}+\frac{-2R}{\hbar}\sum_{n=0}^{n=\infty}a_{n} Im[\frac{\vec{\nabla}^{2n}(R~exp \frac{iS}{\hbar})}{exp \frac{iS}{\hbar}}]=0
 \end{eqnarray}
Therefor the evolution equations for $ \rho $  and $ \vert\psi\vert^{2} $ are different even in a moment if $\rho= \vert\psi\vert^{2} $ is satisfied, this equality will not be established during the next times. However, the probabilistic interpretation of the wave function is not a necessary and we use the essential part of Bohm's theory and without this assumption this theory is meaningful and testable. But experience has shown that in the nonrelativistic limit,  assuming $\rho= \vert\psi\vert^{2} $ is established therefore in some cases, the statistical results of this theory can be extracted even in relativistic regime. For this purpose  we consider a system that its initial state is nonrelativistic so in this circumstance we know $\rho(\vec{r},0)= \vert\psi (\vec{r},0)\vert^{2} $ with good accuracy. Then by placing the system in an external field,  the state of the system will transfer to the relativistic (weak) regime and now we want to to calculate the time evolution of the wave function and location by using equation(\ref{9}) and (\ref{14}).(namely) at any time the particle density $ \rho(\vec{r},t) $ using the continuty equation (\ref{61}) and initial condition $\rho(\vec{r},0)= \vert\psi (\vec{r},0)\vert^{2} $can be calculated. It is clear that in the beginning before the system will transit to the relativistic regime, the relativistic and nonrelativistic evolution equations  are identical with the good accuracy, but after the  transferto the relativistic regime, the path of particles deviate from the path of the non-relativistic case and the condition  $\rho= \vert\psi\vert^{2} $ does not apply and it can be used as an experimental test of this theory. For example we want to consider the scattering of a particle from a potential barrier (e.g.Coulomb potential) and then we can compare the result with experimental results and the predictions of the usual relativistic quantum mechanics. Another example is a particle in a potential well where the volume surrounded by the walls is shrinking. At first, the Compton wavelength of the particle is large compared to the size of the box so we are in the nonrelativistic regime and the quantum equilibrium will be established and when the box size is small, the system is transferred to the relativistic regime and the quantum equilibrium  of the system from thenon-relativistic regime is removed. The results of such an experiment is computed explicitly and a deviation of $ \rho $ from $ \vert\psi\vert^{2} $ is derived as clear criteria for an experimental test of this formalism.
 \section{Relativistic Quantum Equilibrium }
 In this section we want to generalize the assumption of quantum equilibrium to the relativistic case. Although the description of the previous section clearly show that the probability density in the relativistic case deviates from $ \vert \psi\vert^{2} $. The distribution of the relativistic quantum equilibrium can be obtained from these concepts: Non-relativistic limit, Relaxation, Equivariance and Typicality. Non-relativistic limit and  Relaxation are applied to derive an numerical calculation, Equivariance and Typicality are applied to perform analytical expression of relativistic quantum equilibrium. Here we describe a process based on the Equivariance concept and want to answer this equation:
 
 What is the distribution function must be replaced by $ \vert \psi\vert^{2} $ as a quantum equilibrium ?
 
 At first, we investigate the concept of Equivariance more accurately. We consider a general distribution function $\rho^{\psi} $ as a function of the wave function of the systems. Based on the evolution equation of the wave function (for example Schrödinger equation) the probability density changes with time:
 \begin{eqnarray}
 \rho^{\psi_{0}}\longrightarrow\rho^{\psi_{t}}
 \end{eqnarray}
 on the other hand, using the continuity equation, we can examine the time evolution of the probability density:
 \begin{eqnarray}
 \rho^{\psi_{0}}\longrightarrow\rho^{\psi_{0}}_{t}
 \end{eqnarray}
 if these two types of time evolution coincide the probability density is called Equivariancy, namely 
 \begin{eqnarray}
 \rho^{\psi}_{t}=\rho^{\psi_{0}}_{t}
 \end{eqnarray}
  As mentioned before, $\rho^{\psi} $ satisfies in the continuity equation:
  \begin{eqnarray}\label{67}
 \frac{\partial \rho^{\psi} }{\partial t}+\vec{\nabla}\rho^{\psi} . \vec{v}+\rho^{\psi}\vec{\nabla} . \vec{v}=0
 \end{eqnarray}
  where$ \frac{\partial \rho^{\psi} }{\partial t} $ can be calculated in terms of $ \frac{\partial\psi}{\partial t} $ and also the velocity field, $ \vec{v} $ is determined by the equation of particle motion. Therefore the equivariance distribution function can be extracted from \ref{67}. For illustration we look at this process in one dimension, where the continuity equation will be 
 \begin{eqnarray}\label{68}
 \frac{\partial \rho^{\psi} }{\partial t}+\frac{\partial \rho^{\psi} }{\partial x}.v+\rho^{\psi}\frac{\partial v}{\partial x}=0
 \end{eqnarray}
 As a simplest choice we assume that the probability density is a function of wave function and is localized. Therefore by considering the polar form of the wave function :
 \begin{eqnarray}
 \rho^{\psi} =\rho[R,S]
 \end{eqnarray}
 we get
 \begin{eqnarray}\label{70}
 \frac{\partial \rho^{\psi}}{\partial t}=\frac{\partial R}{\partial t}\frac{\partial \rho(R,S)}{\partial R}+\frac{\partial S}{\partial t}\frac{\partial \rho(R,S)}{\partial S}
 \end{eqnarray}
 and then
 \begin{eqnarray}\label{71}
 \frac{\partial \rho^{\psi}}{\partial x}=\frac{\partial R}{\partial x}\frac{\partial \rho(R,S)}{\partial R}+\frac{\partial S}{\partial x}\frac{\partial \rho(R,S)}{\partial S}
 \end{eqnarray}
 By putting equations \ref{70} and \ref{71} in to eq.\ref{68} we have 
 \begin{eqnarray}
 (\frac{\partial R}{\partial t}+\frac{\partial R}{\partial x}v)\frac{\partial \rho(R,S)}{\partial R}+(\frac{\partial S}{\partial t}+\frac{\partial S}{\partial x}v)\frac{\partial \rho(R,S)}{\partial S}+\rho(R,S)\frac{\partial v}{\partial x}=0
 \end{eqnarray}
 $\frac{\partial R }{\partial t}$ and  $\frac{\partial S }{\partial t}$can be calculated from the time evolution of the wave function  in the relativistic level. From above equation,  the relativistic distribution function is obtained, which is equvariante but with the insertion of \ref{14}, (35) and (36) instead of $v$  ,$\frac{\partial R }{\partial t}$ and  $\frac{\partial S }{\partial t}$ , the above equation is more complex and is very difficult to solve  whereas in nonrelativistic level distribution function is obtained as 
 \begin{eqnarray}
 \rho=c_{0}R^{2}
 \end{eqnarray}
 where $c_{0}$  is a constant. Currently an exact solution for the relativistic distribution function is under investigation.
\section{discussion}
In this paper we presente a new relativistic generalization of the Bohmian Mechanics. This formalism implicitly shows that in the relativistic regime, $\rho= \vert\psi\vert^{2} $ is not a valid assumption.
Since the assumption of quantum equilibrium  has a key role in deriving the Born Principle in the causal theory of measurements \cite{Holland},  the measurement results for other quantities will deviate from the  Born principle in the relativistic regime. It is clear that this formalism can not describe the phenomenon of particle creation and annihilation (which mostly occurs at high energy) as  the standard relativistic quantum mechanics, which is also incapable of describing such phenomena. As a result, it is expected that the
formalism presented here is correct only in the limit of weak relativistic regime i.e as long the kinetic energy is smaller than the rest energy and  the probability of particle creation and annihilation is negligible. We describe the proper process to get out of the system from equilibrium quantum that is particularly important because the standard interpretation of the transition from quantum mechanics to quantum field theory is not so clear and the probabilistic wave function and the interpretation of $ | \psi | ^ 2 $ as a probability density, which is accepted in nonrelativistic quantum mechanics, and exactly how and at what stage in the quantum field theory is replaced.

\end{document}